\documentclass[10pt,a4paper,twoside]{article}
\usepackage{epsfig}
\usepackage{baltlat5}
\usepackage{float}
\usepackage{dcolumn}
\begin{document}
\ \
\vspace{-0.5mm}

\setcounter{page}{581}
\vspace{-2mm}

\titlehead{Baltic Astronomy, vol.\ts 15, 581--599, 2006.}

\titleb{RADIATIVE TRANSFER PROBLEM IN DUSTY GALAXIES:\\
PHOTOMETRIC EFFECTS OF DUSTY SPIRAL ARMS}

\begin{authorl}
\authorb{D.~Semionov}{1},
\authorb{K.~Kodaira}{2},
\authorb{R.~Stonkut{\. e}}{1,3} and
\authorb{V.~Vansevi{\v c}ius}{1}
\end{authorl}

\vbox{
\begin{addressl}
\addressb{1}{Institute of Physics, Savanori{\c u} 231,
Vilnius LT-02300, Lithuania}

\addressb{2}{The Graduate University for Advanced Studies (SOKENDAI),\\
Shonan Village, Hayama, Kanagawa 240-0193, Japan}

\addressb{3}{Vilnius University Observatory, \v{C}iurlionio 29, Vilnius LT-03100, Lithuania}
\end{addressl}
}
\vskip-1mm

\submitb{Received 2006 November 17; revised and accepted 2006 December
11}

\begin{summary} We study the effects of dusty spiral arms on the
photometric properties of disk galaxies using a series of 2D radiative
transfer models, approximating the arms with axially symmetrical rings.
We find that dusty arms, as well as dusty disks, have a significant
influence on the aperture photometry and surface brightness profiles
altering colors of model galaxies.  We suggest that, in addition to the
conventionally modeled diffuse absorbing layers or disks, the dusty arms
should be taken into account in spiral galaxy extinction studies.
\end{summary}

\begin{keywords}
ISM: dust, extinction -- galaxies: spiral --  galaxies: ISM -- galaxies:
photometry
\end{keywords}

\resthead{Photometric effects of dusty spiral arms}{D. Semionov, K.
Kodaira, R. Stonkut\.e, V. Vansevi\v cius}

\sectionb{1}{INTRODUCTION}

Since the pioneering work on the extinction effects in galaxies by
Holmberg (1958) through the recent papers (Disney, Davies \& Phillips
1989; Kodaira, Doi \& Shimasaku 1992; Corradi, Beckman \& Simonneau 1996
and others referred therein) it has been assumed for modeling that the
absorbing matter is smoothly distributed in the geometry of layers
parallel to the stellar disk.  In particular, various inclination
effects have been interpreted using the absorption-layer models, leading
to partly contradicting inferences about the opaqueness of galaxy disks
(see Valentijin 1990; Burstein, Haynes \& Faber 1991; Huizinga \& van
Albada 1992; Davies \& Burstein 1995).  Conclusions of these statistical
studies strongly depend on the studied galaxy samples which show a wide
variety even among galaxies of a given morphological type.  Kodaira \&
Yamashita (1996) analyzed edge-on Scd (T\,=\,6) galaxies NGC\,3556 and
NGC\,4244 and pointed out the contrasting nature between the two
examples.  The former is rich in dark-lane features and has a high index
of $B_T-m{\rm FIR}=+1.48$, while the latter scarcely shows dark lanes
and has a low index of $B_T-m{\rm FIR}=-0.3$.  By applying exponential
layer models for the stellar disk and the absorbing matter, the authors
found that the central dust layer of NGC\,3556 has an outer cut-off
radius smaller than the stellar disk extent and opaque patches which may
partly be due to spiral arms.  We suspect that in NGC\,3556 the dark
arms may play a significant role in the total extinction in addition to
the diffuse absorbing component.

Normally, in face-on galaxy images we ignore the spiral dark lanes
because of small cross-section on the galactic plane and of the assumed
small height above the plane.  However, Sofue (1987) and Sofue,
Wakamatsu \& Malin (1994) suggested that the spiral dark lanes in
general may have vertical structures which reach high above the stellar
disk.  They investigated NGC\,253 (Sc) and NGC\,7331 (Sb) and found
abundant dark structures rising from the spiral dark lanes up to a few
kpc above the disk plane.  By examining images of edge-on galaxies in
the Hubble Atlas, similar structures are easily found for NGC\,891 (Sb)
and NGC\,4631 (Sc).  The four galaxies mentioned above all have a high
index of $B_T-m{\rm FIR}>1.5$ according to RC3 (de Vaucouleurs \etal\
1991), indicating enhanced star-formation activity (see Tomita, Tomita
\& Saito 1996 and references therein).  Bosma \etal\ (1992) made
velocity-profile analysis of NGC\,891 using H\,I, CO and H$\alpha$
lines, to find that the inner part of this galaxy is optically thick
while the outer part is optically thin when seen at face-on.  Huizinga
\& van Albada (1992) made the same conclusion for Sc galaxies in
general.  We are reminded that spiral arms often occupy a larger
fraction of space in the inner part of disk galaxies than in the outer
part.  Block, Elmegreen \& Wainscoat (1995) studied NGC\,2841 in near IR
to reveal underlying spiral dark lanes behind the amorphous smooth
optical images.

These findings support the suspicion that the dark lanes may play a
significant role in internal extinction of some disk galaxies.  Corradi
\etal\ (1996) in their photometric models of disk galaxies considered
differences of stellar populations between the arm and the inter-arm
regions, however, they adopted the layer geometry for dust distribution.
Wainscoat \etal\ (1992) adopted a logarithmic spiral pattern in
constructing a galaxy model for distribution of young massive stars, but
an exponential layer model for absorbing matter.  Misiriotis \etal\
(2000) have found that a double-exponential disk provides a very good
fit for dusty spiral galaxies seen edge-on.  Steep gradients and high
contrasts between arm and inter-arm regions, apparent in face-on views
of their model, possessing values of $R_{\rm eff}$ for dust and star
distributions and up to 1/3 of the dust locked in spiral arms, are
largely averaged over at high inclinations.  Therefore, it is important
to provide means to model the changes of the observed properties of
galaxies in transition between edge-on and face-on orientations.

Strong color gradients, as well as large variations in disk flatness
(scale-height to scale-length ratio, $Z_{\rm eff}/R_{\rm eff}$), are
observed in most late-type galaxies.  Furthermore, many objects display
variable color gradients or their absence (``flat'' color profiles),
which can be related to their evolution stage or internal extinction
properties (Reshetnikov, Dettmar \& Combes 2003).  Therefore, in order
to establish validity limits of population synthesis model employment
for computation of the radial color profiles, it is important to
estimate the effects of dusty spiral arms on model galaxy color
gradients (color excess profiles).

However, it should be stressed that a spiral pattern in galaxies is
traced not only in the distribution of the interstellar medium, but also
in the young stellar populations, i.e., the stellar arms.  Their
presence can significantly alter photometric properties of galaxies and,
in order to compare model predictions with observations for local well
resolved and distant semi-resolved or unresolved galaxies, the modeling
procedure must take all these features into account.  In this study we
focus on the additional extinction effects, produced by dusty arms, on
colors of stellar disk populations in spiral galaxies.  Our aim is to
estimate the possible range of magnitude of these effects, thus we
choose rather extreme cases of the dusty disks and the dusty arms.  Due
to these simplifications, direct application of our results for
interpretation of the observations is strongly restricted to some
specific cases, e.g., for galaxies in which energy, radiated by young
stellar populations concentrated in spiral arms, does not dominate the
total energy radiated by the disk stellar populations.  However, the
presented results can serve as a guide for the spiral galaxy extinction
estimate in general.

In this paper we present the radiative transfer code and galaxy models
employed (section 2), examine the galaxy inclination and dusty spiral
arm effects on the surface brightness and color excess profiles of
bulge-less (late type, Sc-Sd) spiral galaxies, relevant to the
photometric galaxy survey and number count interpretation (section 3),
and give brief conclusions.

\sectionb{2}{RADIATIVE TRANSFER CODE AND MODEL GALAXIES}

Realistic evaluation of dusty spiral arm influence on the observed
photometric properties of disk galaxies has to be performed fully
accounting for multiple light scattering events and presence of the
dusty disk, possessing diffuse distribution of interstellar dust.  To
test these effects the radiative transfer problem has been solved for
the eight model galaxies M1--M8 (Table~1), representing several star and
dust distributions, using the Galactic Fog Engine code (Semionov \&
Vansevi{\v c}ius 2005c).

This code realizes an iterative ray-tracing algorithm in 2D
axi-symmetrical geometry, allowing for a flexible definition of stellar
and dust content distribution, producing intensity maps of the model
under arbitrary inclination at a given set of wavelengths/passbands.
Computations are performed within a cylinder, which is subdivided into a
set of layers of concentric internally homogeneous rings of varying
vertical and radial extent.  This cylinder is then sampled using a set
of adaptively chosen directions (rays), along which one-dimensional
radiative transfer problems are solved.  The number and placement of
rays depends on stellar light and dust distribution.  In the course of
the first iteration the initial radiative energy of the system is
separated into escaped, that directly reaches an external observer,
absorbed by dust grains, and non-isotropically scattered (Semionov \&
Vansevi{\v c}ius 2005a) radiative energies.  The solution is then
repeated, substituting scattered radiative energy as the initial
distribution for the next iteration, accumulating resulting escaped and
absorbed radiative energies, until certain convergence criteria are met
-- either a fixed number of iterations is reached, or the remaining
scattered radiative energy drops below the specified threshold.  The
Galactic Fog Engine has been carefully tested against several previously
published radiative transfer codes (Witt, Thornson \& Capuano 1992;
Ferrara \etal\ 1999), and the results were found to be deviating by less
than a few percent from the published values (Semionov \& Vansevi{\v
c}ius 2002).

Each model (see Table~1) was represented by a cylinder consisting of 35
concentric rings, subdivided into 49 layers, with radial and vertical
extent of 6 and 2.5 stellar disk scale-lengths, $R_{\rm eff}$,
respectively.  The radiative transfer was evaluated using seven
ray-tracing iterations, of which the first three were computed directly
and the remaining four -- using iteration scaling approximation
(Semionov \& Vansevi{\v c}ius 2005b).  The radiative energy defect and
the radiative energy remaining to be scattered after the last iteration
were below 1\% of the total emitted radiative energy in all cases.

\begin{figure}[H]
\centerline{\psfig{figure=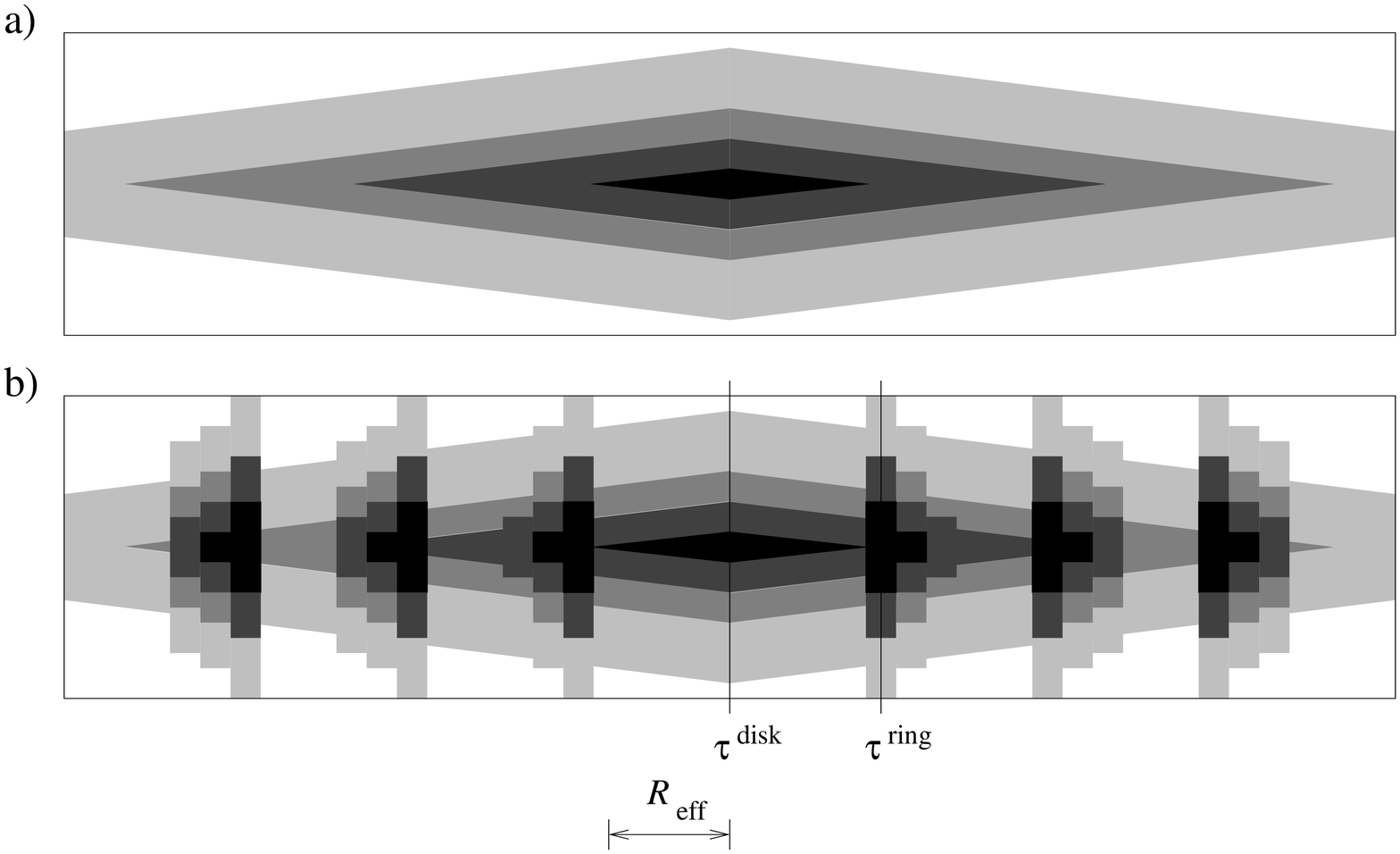,width=120mm,angle=0,clip=}}
\captionb{1}{Diametral cross-section of the interstellar dust
distributions in model galaxies. Panels (a) and (b) show the
schematic cross-section of models with dusty disk and with both, a
dusty disk and dusty rings, respectively. Shading represents
varying relative dust mass density. Marked are the scale-length of
stellar disk, $R_{\rm eff}$, and the radial positions at which
optical depths $\tau^{\rm disk}$ and $\tau^{\rm ring}$ are
measured.}
\end{figure}
\vskip1mm

All models have identical stellar populations, distributed according
to a double-exponential law:
\begin{equation}
\rho(R,Z) = \rho_0 \exp \left( -{R\over R_{\rm eff}} -{\left| Z \right| \over Z_{\rm eff}}\right),
\end{equation}
\noindent where $R_{\rm eff}$ and $Z_{\rm eff}$ are the
scale-length and scale-height, respectively. To simulate the
effect of dust residing in the galaxy disk and arms, extinction in
the model galaxy was distributed in the dusty disk and dusty arms.
Model galaxy schemes are shown in Figure~1 as $(R,Z)$ diametral
cross-sections of the models with gray-scale tones indicating the
relative density of the interstellar dust. Dusty disk extinction
distribution follows a double-exponential law with varying
scale-length, $R_{\rm eff}^{\rm d}$, and scale-height, $Z_{\rm
eff}^{\rm d}$, for different models. Central optical depth in
$V$-band (measured to the central plane of the model galaxy in the
direction perpendicular to it) is set to $\tau_V^{\rm disk}=5$ for
all models.

The dusty arm (``ring'') extinction distribution in the model
galaxy is represented by three identical dusty rings situated at
radial distances $R=1.25$, 2.5 and $3.75R_{\rm eff}$ from the
model axis. Each ring has a radial extent of $0.45R_{\rm eff}$ and
consists of three radial zones of equal width of $0.15R_{\rm
eff}$. The total optical depth of the zones in the $V$-band (measured
to the central plane of the model galaxy in the direction
perpendicular to it) is set to: $\tau_V^{\rm ring}=5$ (inner
zone), 2 (middle zone) and 1 (outer zone), respectively (see
Figure~1). Vertical dust distribution within each zone is
described by the one-dimensional exponential law
\begin{equation}
\rho(R,Z) = \rho_0 (R) \exp \left( -{\left| Z \right| \over Z_{\rm eff}^{\rm d}}\right),
\end{equation}
\noindent assuming identical scale-heights, $Z_{\rm eff}^{\rm d}$,
for the dusty rings and the dusty disk for each model.

Both, disk- and ring-distributed dust, have the same optical properties,
computed using the Laor \& Draine (1993) model, approximating Milky Way
galaxy type extinction, using a mixture of graphite and silicate grains
in the proportion of 0.47 to 0.53 (Mathis, Rumpl \& Nordsiek 1977) and
grain size distribution following the power law $a^{-3.5}$ possessing
lower and upper cut-off radii $a_{\rm min}=0.005$ $\mu$m and $a_{\rm
max}=0.25$ $\mu$m.

The geometric parameters of star and dust distributions for the
model galaxies M1--M8, discussed in this work, are presented in
Table~1. All models have identical stellar
populations, described using the scale-length, $R_{\rm eff}$, and
-height, $Z_{\rm eff}$. To simulate the commonly used relative
star-to-dust test distributions the respective scale-length,
$R_{\rm eff}^{\rm d}$, and -height, $Z_{\rm eff}^{\rm d}$, of the
dusty disk were varied, producing ``standard'' ($R_{\rm eff}^{\rm
d} = R_{\rm eff}$ and
 $Z_{\rm eff}^{\rm d} = Z_{\rm eff}$; M1 and M5),
``thin''
($R_{\rm eff}^{\rm d} = R_{\rm eff}$ and
 $Z_{\rm eff}^{\rm d} = 0.5 Z_{\rm eff}$; M2 and M6),
``thick''
($R_{\rm eff}^{\rm d} = R_{\rm eff}$ and
 $Z_{\rm eff}^{\rm d} = 2.0 Z_{\rm eff}$; M3 and M7), and
``wide''
($R_{\rm eff}^{\rm d} = 2.0 R_{\rm eff}$ and
 $Z_{\rm eff}^{\rm d} = Z_{\rm eff}$; M4 and M8) models.
Models M1--M4 represent pure disks, similar to bulge-less SxxMExx
models in Ferrara \etal\ (1999), with dust extinction present only
in a form of a double-exponential law with constant central optical
depth, $\tau_V^{\rm disk}$, for all cases, while models M5--M8
additionally contain the dusty arms with the same maximum optical
depth, $\tau_V^{\rm ring}$, for all rings. The resulting total dust
mass, $M_{\rm tot}^{\rm d}$, in Solar mass units, required to
produce this amount of opacity, is given in Table~1.

\begin{center}
\vbox{\footnotesize
\begin{tabular}{cccccccc}
\multicolumn{8}{c}{\parbox{8cm}{\baselineskip=8pt ~~~~{\smallbf
Table~1.}{\small\ Parameters of the studied model galaxies.
$R_{\rm eff}$ and $Z_{\rm eff}$ are the scale-length and -height
of double-exponential stellar disk, while $R_{\rm eff}^{\rm d}$
and $Z_{\rm eff}^{\rm d}$ mark the respective quantities for the
dusty disk. $\tau_V^{\rm disk}$ and $\tau_V^{\rm ring}$ are the
maximum optical depths (measured to the central plane of the model
galaxy in the direction perpendicular to it) for the dusty disk
and the dusty arms, respectively, as shown in Figure~1. $M_{\rm
tot}^{\rm d}$
is the total dust mass (disk+arms) present in the model, in \msun\ units.}}}\\
\tablerule Model & $R_{\rm eff}$ & $Z_{\rm eff}$ & $R_{\rm
eff}^{\rm d}$ & $Z_{\rm eff}^{\rm d}$ &
$\tau_V^{\rm disk}$ & $\tau_V^{\rm ring}$ & $\log M_{\rm tot}^{\rm d}$ \\
\tablerule
M1 & 2.0 & 0.2 & 2.0 & 0.2 & 5 & 0 & 7.55\\
M2 & 2.0 & 0.2 & 2.0 & 0.1 & 5 & 0 & 7.55\\
M3 & 2.0 & 0.2 & 2.0 & 0.4 & 5 & 0 & 7.55\\
M4 & 2.0 & 0.2 & 4.0 & 0.2 & 5 & 0 & 8.06\\
M5 & 2.0 & 0.2 & 2.0 & 0.2 & 5 & 5 & 8.01\\
M6 & 2.0 & 0.2 & 2.0 & 0.1 & 5 & 5 & 8.01\\
M7 & 2.0 & 0.2 & 2.0 & 0.4 & 5 & 5 & 8.01\\
M8 & 2.0 & 0.2 & 4.0 & 0.2 & 5 & 5 & 8.25\\
\tablerule
\end{tabular}
}\label{tab:first}
\end{center}

For all models we have computed 1000$\times$1000 pixel images with a
scale of $0.0125 R_{\rm eff}$ per pixel at the wavelengths corresponding
to the $B$, $V$ and $K$-bands for model galaxy inclinations between
0\degr\ and 90\degr\ with a step of 5\degr.  We have performed surface
photometry of these images using concentric apertures with ellipticity
determined from the model inclination, the apertures being centered on
the geometric center of the model galaxy image.  Magnitudes and color
excesses in growing apertures, differential azimuthally-averaged surface
brightness and color excess profiles, as well as various cross-sections
of the model galaxies, were derived and analyzed.

\sectionb{3}{RESULTS}

\subsectionb{3.1}{Minor axis cross-sections}

From the synthetic images, described in the previous section, we have
extracted photometric minor axis cross-sections integrated within
one-pixel-wide ($0.0125 R_{\rm eff}$) image strip, passing through the
center of the galaxy, for each considered model.  The results are
presented in Figure~2.  Each panel shows surface brightness in $V$-band
and $E_{B-V}$ and $E_{V-K}$ color excess cross-sections determined for
``standard'' (M5, solid line), ``thin'' (M6, dotted line), ``thick''
(M7, dashed line) and ``wide'' (M8, dash-dotted line) models.
Cross-section data are plotted against distance from the model galaxy
image major axis, measured in $R_{\rm eff}$, for inclination angles of
0\degr, 45\degr, 70\degr, 80\degr, 85\degr\ and 90\degr\ (panels
arranged in left-to-right direction as indicated in the topmost panel
row).
\begin{figure}[H]
\centerline{\psfig{figure=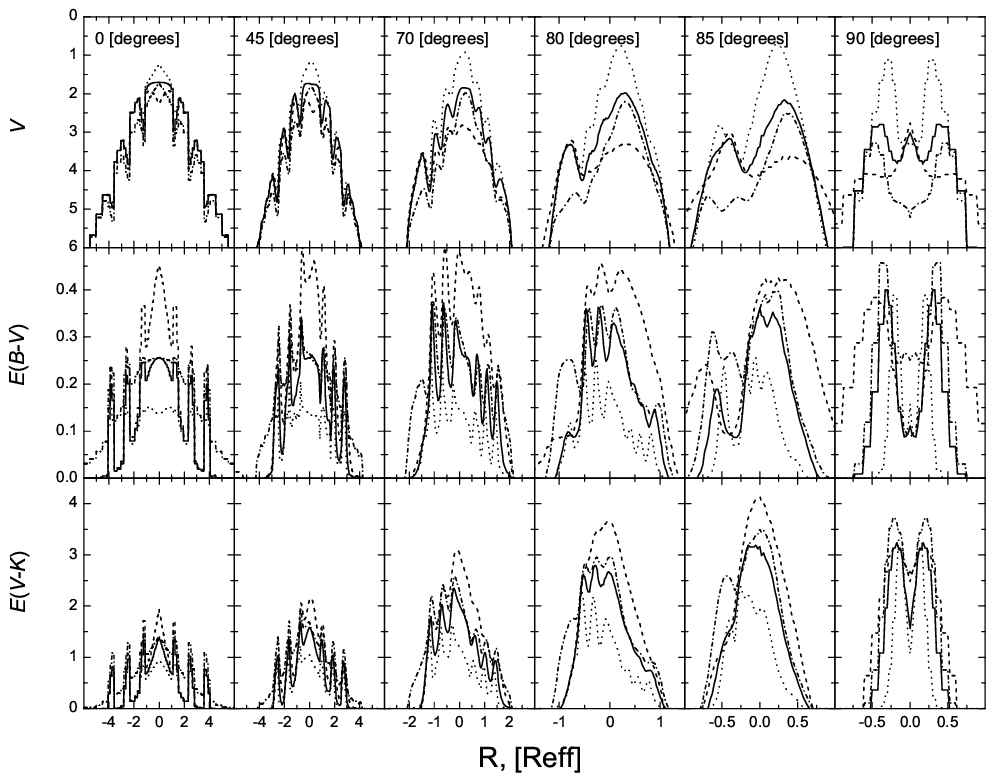,width=120mm,angle=0,clip=}}
\captionb{2}{Inclination effect on model galaxy image minor axis
cross-sections. Each panel shows surface brightness in $V$-band
and $E_{B-V}$ and $E_{V-K}$ color excess cross-sections determined
for ``standard'', ``thin'', ``thick'' and ``wide'' (solid, dotted,
dashed and dash-dotted lines, correspondingly) models plotted
against distance from the model galaxy image major axis, measured
in $R_{\rm eff}$. Panels are arranged by inclination in the
left-to-right direction as indicated in the topmost panels.}
\end{figure}

As can be seen in Figure~2, the minor axis cross-sections of the model
galaxies clearly show the dusty arms (individual marks of the arms start
to merge at a critical angle of about 70\degr) as well as increasing
asymmetry with increasing inclination.  $V$-band surface brightness
cross-sections, presented in top panels of Figure~2, show increasing
separation between different models with increasing inclination.
Similar trends are also observed for $E_{B-V}$ and $E_{V-K}$ color
excesses.

Characteristic asymmetries of high inclination (60\degr\ to 85\degr)
cross-sections allow one to classify galaxies qualitatively possessing
different dust distributions with respect to stellar disk into at least
three broad groups.  Cross-sections at lower inclinations are less
sensitive and more degenerate to the dusty disk parameters and hardly
could be applied for the internal extinction distribution analysis in
spiral galaxies.  The transition between two characteristic inclination
angle ranges -- low- (0\degr\ -- 70\degr) and high-inclination (70\degr\
-- 90\degr) -- is determined by ``inter-arm shadowing angle'', which for
the model galaxies, analyzed in this study, is approximately 70\degr.
Under this inclination the effective extinction zone (defined as the
dusty arm zone with the projected optical depth of $\tau_V>1$) of the
dusty arms significantly shadows the model's center and the inter-arm
regions from direct observation.

\subsectionb{3.2}{Azimuthally averaged profiles}

Azimuthally averaged surface brightness and color excess profiles
were computed using a set of elliptical apertures, centered on the
geometrical center of the model galaxy with aperture axis ratio,
$b/a$, determined from the model inclination angle, $i$, using the
following relation
\begin{equation}
{b \over a} = \cos i + {\chi^2 \sin^2 i \over \cos i + \chi \sin i},
\end{equation}
\noindent where $\chi=Z_{\rm eff}/R_{\rm eff}$. Due to the annular
arrangement of dusty arms and neglected effects of dust clumpiness
in our models the azimuthally averaged profiles computed for model
inclination of 90\degr\ are less realistic, since the opaque outer
dusty arm cuts out the light from the central part of the model
too effectively. However, the derived total photometric parameters
of the model galaxy are expected to be correct (Semionov \&
Vansevi{\v c}ius 2002).

Inclination effects on the model galaxy surface brightness and color
excess profiles for models with both, a dusty disk and dusty arms
(models M5--M8), are shown in Figure~3.  Panels, arranged by the model
galaxy inclination angle in the left-to-right direction, as indicated in
the topmost row, show $V$-band surface brightness (topmost panel row)
and differential $E_{B-V}$ and $E_{V-K}$ color excess (second and third
panel rows, respectively) profiles for ``standard'' (M5, solid line),
``thin'' (M6, dotted line), ``thick'' (M7, dashed line), and ``wide''
(M8, dash-dotted line) models as a function of radial distance given in
units of $R_{\rm eff}$.

As can be seen in Figure~3, in low-inclination (up to $\sim$70\degr)
cases the azimuthally averaged $V$-band surface brightness radial
profiles show presence of all three dusty arms for all models, while at
the high inclinations only ``thin'' model profile shows any coherent
structure.  It is worth noting, that $E_{B-V}$ radial profiles retain
information about the number and position of dusty arms up to higher
inclinations than $E_{V-K}$ (cf. the relative structure of color excess
profiles at inclinations of 45\degr\ and 70\degr).  Also a significant
flattening of surface brightness profiles at high inclinations (80\degr\
and above) is noticeable, while color excess profiles still show the
radial gradients comparable or exceeding those observed in
low-inclination model galaxies.

\vbox{
\centerline{\psfig{figure=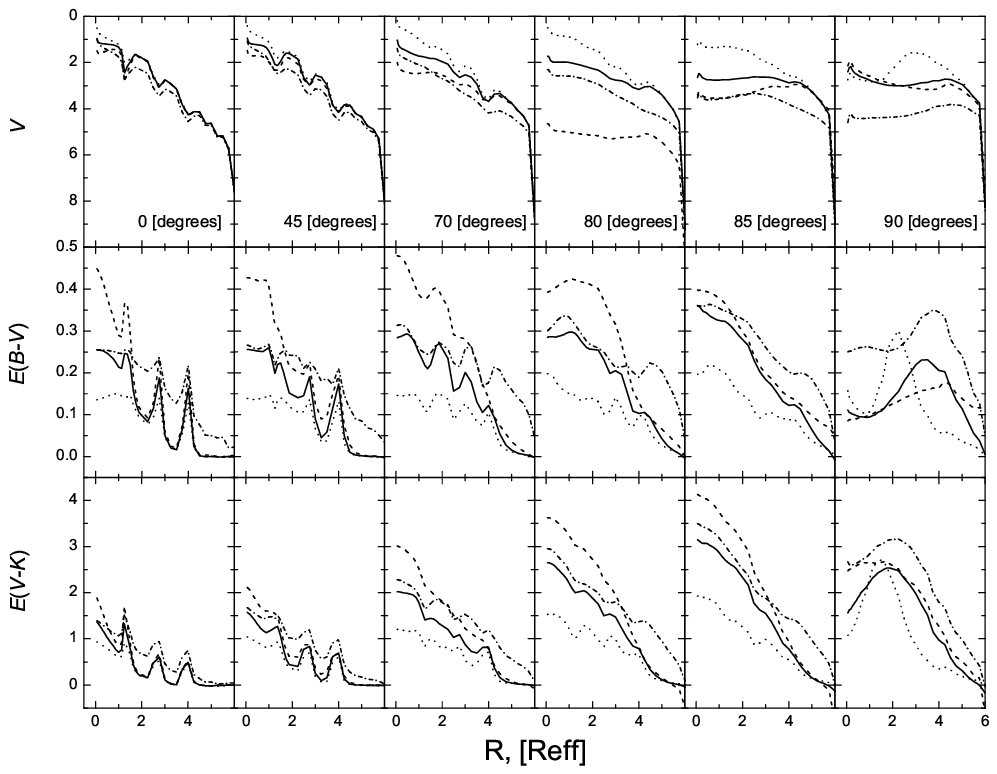,width=120mm,angle=0,clip=}}
\captionb{3}{Inclination effects on the model galaxy surface
brightness and color excess profiles. Panels show surface
brightness in $V$-band and $E_{B-V}$ and $E_{V-K}$ color excess
profiles for ``standard'', ``thin'', ``thick'' and ``wide''
models (solid, dotted, dashed and dash-dotted lines,
respectively) measured in rings of increasing radius, given in
units of $R_{\rm eff}$. Panels are arranged by model galaxy
inclination in left-to-right direction as indicated in topmost
panels.}
}
\vspace{3mm}

In order to determine the influence of the dusty arms on photometry
performed in elliptical apertures and azimuthally averaged radial
profiles we directly compared $E_{B-V}$ color excess and color excess
ratio $E_{V-K}/E_{B-V}$ derived from model galaxies possessing only a
dusty disk (shown with a solid line) with models having both, a dusty
disk and dusty arms (shown with filled circles connected by a solid
line).  Figures~4 and 5 show $E_{B-V}$ and $E_{V-K}/E_{B-V}$ quantities
in growing apertures as a function of aperture major axis, expressed in
$R_{\rm eff}$ units, while Figures~6 and 7 show the differential radial
profiles of the respective quantities.

As can be seen in Figure~4, the dependence of the $E_{B-V}$ color excess
on the aperture size is determined primarily by the inter-relation of
dusty disk scale-height to scale-length ratio, $\chi^{\rm d}=Z_{\rm
eff}^{\rm d}/R_{\rm eff}^{\rm d}$, with the same parameter of the
stellar disk, $\chi$.  This results in the ``wide'' model color excess
behaving almost identical to the ``thin'' model.  Furthermore, the
difference between models with and without dusty arms depends on the
value of $\chi^{\rm d}$, being negligible for ``wide'' and ``thin''
models and strongly increasing for ``normal'' and ``thick'' models.
This behavior is observed for all inclinations up to nearly edge-on
cases (inclination of 85\degr).  Under the assumption that the dust
distribution is generally more concentrated towards the central plane
than the star distribution ~($\chi^{\rm d} < \chi$), we conclude
that the\break

\vbox{
\centerline{\psfig{figure=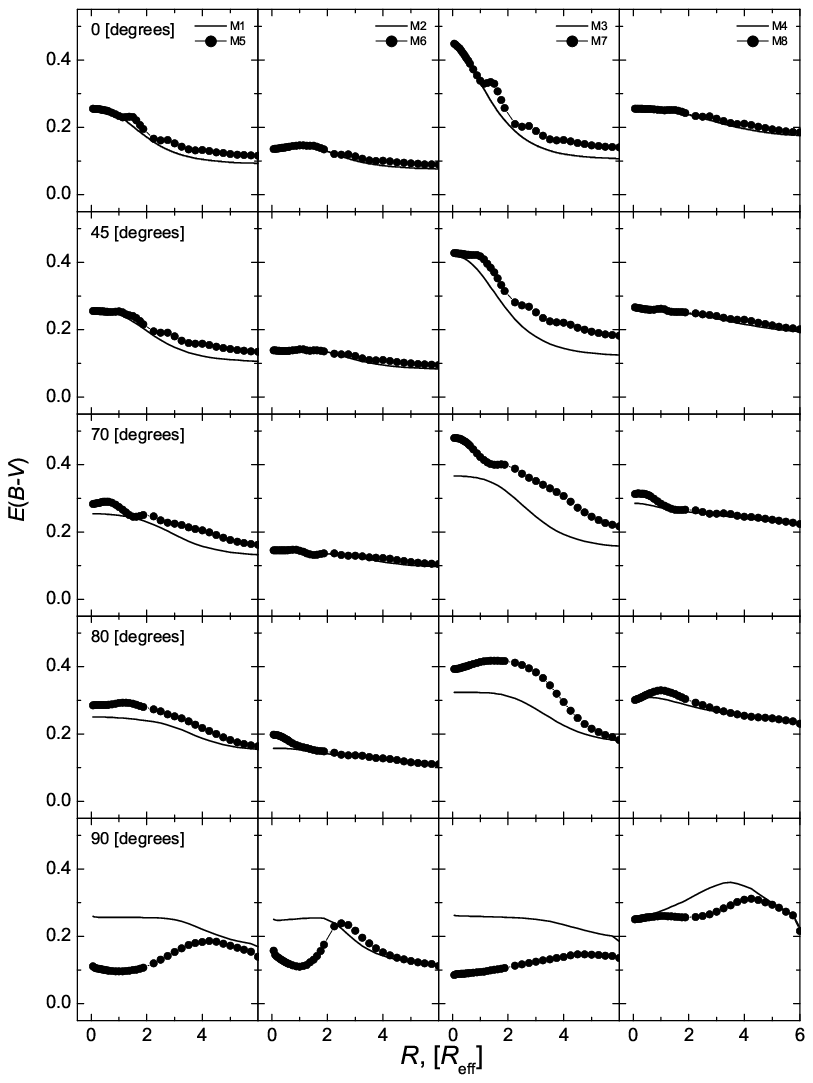,width=120mm,angle=0,clip=}}
\vspace{2mm}
\captionb{4}{ $E_{B-V}$ color excess measured in a sequence of
growing apertures versus radius given in units of $R_{\rm eff}$.
Panels are arranged according to inclination angle (0\degr,
45\degr, 70\degr, 80\degr\ and 90\degr) and dust-to-star relative
distribution (``standard'', ``thin'', ``thick'' and ``wide'') in
rows and columns, respectively. Solid lines show results obtained
for the model galaxies with a dusty disk, solid lines with filled
circles show results obtained for the model galaxies with both, a
dusty disk and dusty arms.}
}

\vbox{
\centerline{\psfig{figure=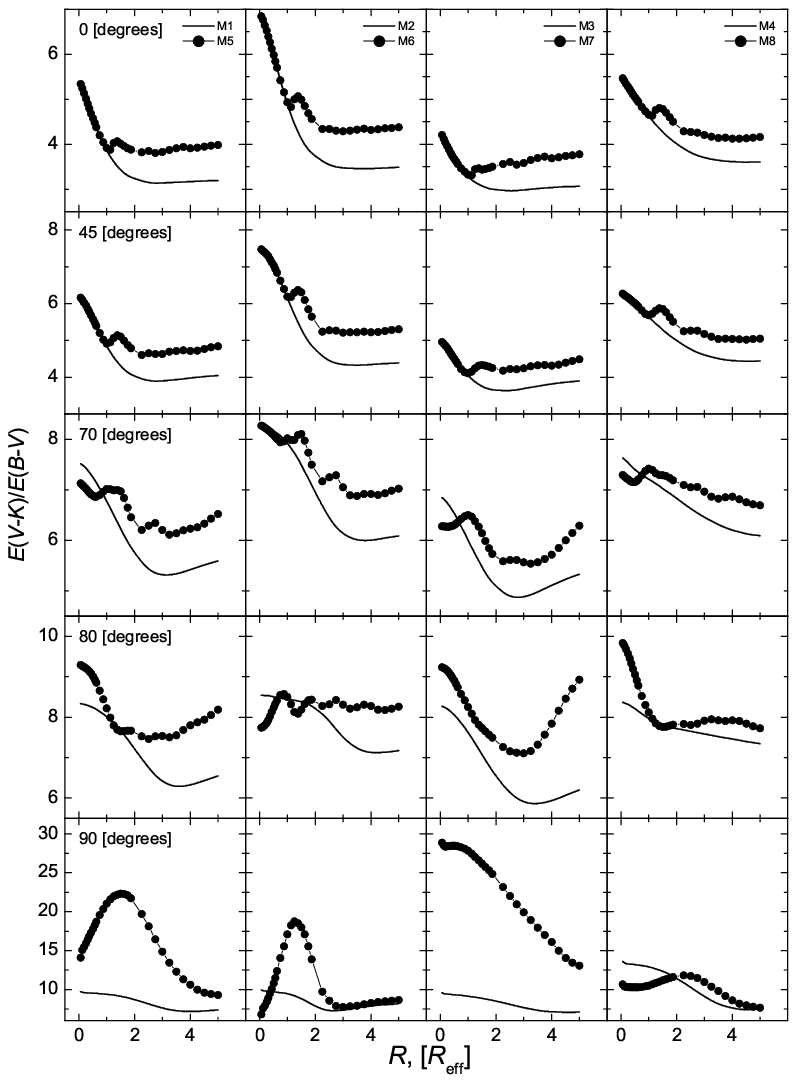,width=120mm,angle=0,clip=}}
\captionb{5}{The same as in Figure~4, but for the color excess ratio
$E_{V-K}/E_{B-V}$.}
}

\vspace{3mm}
\noindent
$E_{B-V}$ color excess value, measured within aperture extending more
than $2R_{\rm eff}$ from the galaxy image center (approximately the
distance completely covering the first dusty arm in our model galaxies),
is insensitive to the presence of dusty arms (as is commonly assumed on
the grounds of small relative cross-section of the dusty arms) for all
except edge-on galaxies.

\newpage
\vbox{
The dependence of the color excess ratio $E_{V-K}/E_{B-V}$ on aperture
size, shown in Figure~5, however, is significantly more complicated.  In
low-inclination cases (up to 70\degr) for the apertures, covering more
than half of the total model galaxy radius ($>$\,$3R_{\rm eff}$), the
introduction of dusty arms (and thus an increase in total extinction
and, therefore, in expected dust mass) leads to the increase in
$E_{V-K}/E_{B-V}$.  The dependence of the color excess ratio on aperture
size beyond $3R_{\rm eff}$ for models with and without dusty arms is
very similar, thus it might be possible that the introduction of dusty
arms affects $E_{V-K}/E_{B-V}$ in a manner identical or similar to an
increase of the dusty disk's total extinction.  However, this statement
requires further investigation.  It should also be noted, that color
excess ratio $E_{V-K}/E_{B-V}$, measured within small apertures
($\sim$\,$R_{\rm eff}$ or less) does not provide a good estimate of
color excess ratio for the entire model galaxy and is strongly dependent
on dust distribution geometry, particularly for high-inclination cases.

Significant is the fact, that in the case of aperture photometry for
low-inclination models (inclination of up to 50\degr, Figures~4 and 5)
the respective values remain nearly constant for the large extent of
the profile, thus allowing the determination of total dust content and
scale-length of dusty disk using aperture photometry.

Radial profiles of $E_{B-V}$ color excess and color excess ratio
$E_{V-K}/E_{B-V}$ (Figures~6 and 7, respectively), derived from
azimuthally averaged surface brightness profiles, show that for all
models, seen under low inclination (below 70\degr), dusty arm effects
are well localized.  The central, outer and inter-arm region profiles of
the model galaxies with both, a dusty disk and dusty arms, are very
similar to the profiles computed for the models with a dusty disk only.
The presence of the dusty arms is clearly seen in all radial color
excess profiles of low-inclination models and their number and position
(radial distance) can be easily determined.  For higher inclination
models (80\degr\ -- 90\degr) the radial color excess profiles show no
significant effect due to presence of dusty arms as the individual dusty
arm profiles become overlapped and indistinguishable.

The photometric effect of the dusty arms significantly depends on
optical density (opacity) difference between the dusty arms and the
dusty disk (``arm contrast'').  The dusty arms are the least prominent
in the ``wide'' model and the most prominent in the ``thick'' model.  As
is obvious from geometric considerations, the ``thick'' model color
excess profile also displays the prominent effects of overlapping dusty
arms under lower inclinations than other models, the effect already
being significant at 70\degr\ inclination.

Color excess ratio $E_{V-K}/E_{B-V}$ profiles (Figure~7) behave
similarly to the above discussed $E_{B-V}$ color excess profiles,
additionally showing presence of dusty arm-scattered light in the
outermost parts of the models under low inclinations.  However, in this
case the most prominent effects of dusty arms are seen in the ``thin''
model profiles.

\subsectionb{3.3}{Effective model galaxy extinction and color excesses}

In the following we discuss the inclination dependence of the total
extinction in $V$-band, $A_V$ and $A_V/E_{B-V}$ as well as $A_V/E_{V-K}$
ratios.  The effect of multiple photon scattering from the dusty arms
can be tested using the innermost part of the model, enclosed within the
first dusty arm, particularly when seen under low inclination, while the
direct shadowing of the model central region by the dusty arms is
negligible.  Furthermore, due to selection effects, only the brightest
parts of the distant galaxies are usually observable.

}

\begin{figure}[H]
\centerline{\psfig{figure=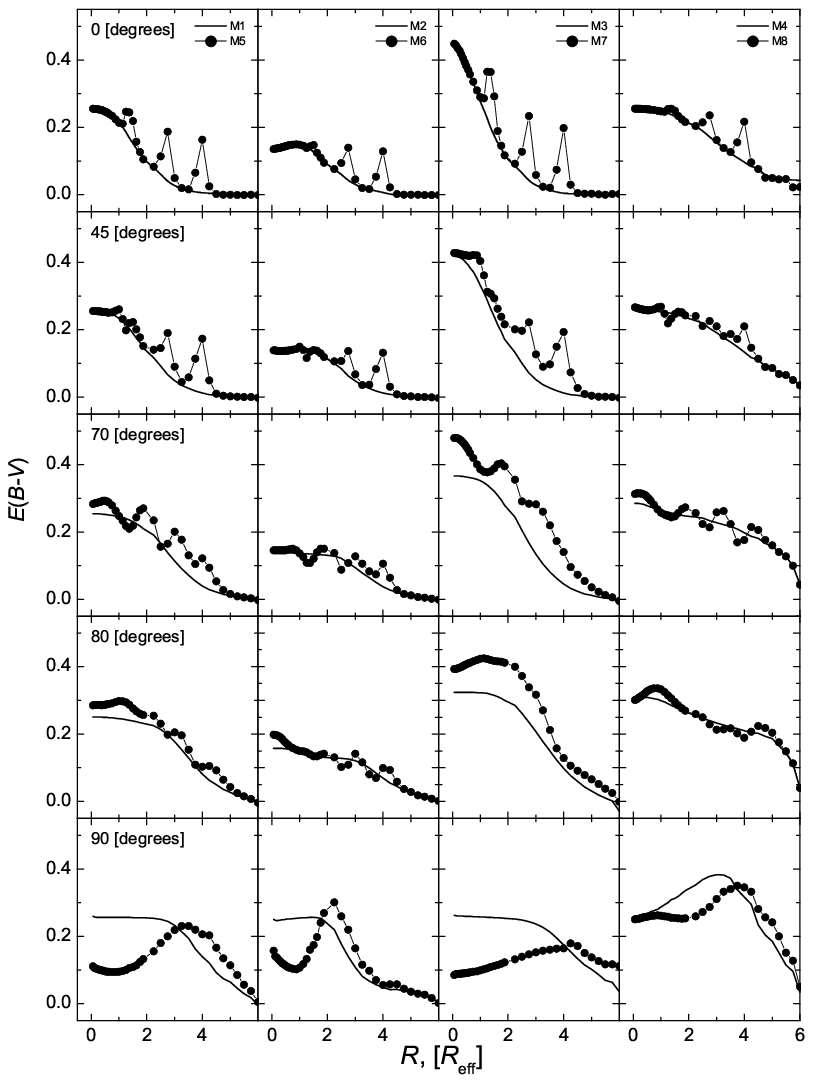,width=125mm,angle=0,clip=}}
\captionb{6}{ $E_{B-V}$ color excess measured in a sequence of
elliptical annuli. Panels are arranged according to inclination
angle (0\degr, 45\degr, 70\degr, 80\degr\ and 90\degr) and
dust-to-star relative distribution (``standard'', ``thin'',
``thick'' and `wide'') in rows and columns, respectively. Solid
lines show results obtained for the model galaxies with a dusty
disk, solid lines with filled circles show results obtained for
the model galaxies with both, a dusty disk and dusty arms.}
\end{figure}

\begin{figure}[H]
\centerline{\psfig{figure=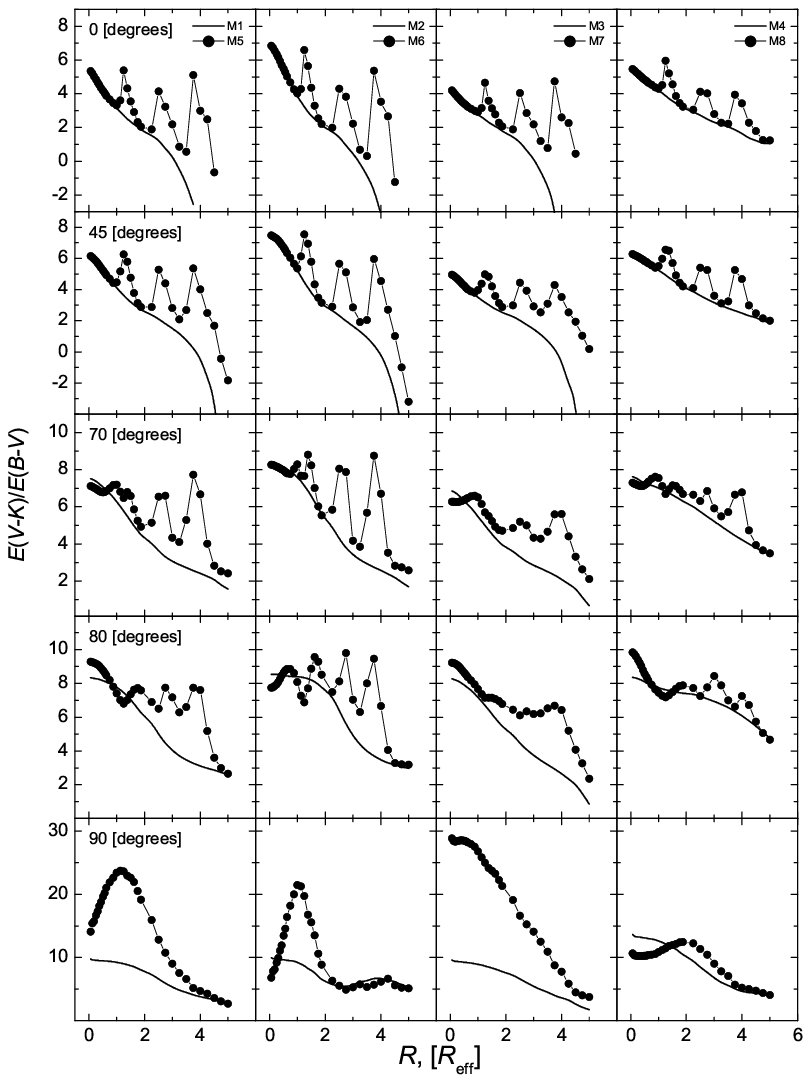,width=120mm,angle=0,clip=}}
\captionb{7}{The same as in Figure~6, but for the color excess ratio
$E_{V-K}/E_{B-V}$.}
\end{figure}

Therefore, we
present two sets of results:  the photometry parameters obtained for
aperture with semi-major axis radius of $1R_{\rm eff}$ (the central
model galaxy region fully enclosed within the innermost dusty arm) and
for aperture with semi-major axis radius of $6R_{\rm eff}$ (the entire
model galaxy).

As can be seen in the upper row of Figure~8, in most low inclination (up
to 60\degr) cases the presence of dusty arms has no significant impact
on the extinction in $V$-band in the central part of the model galaxy.
The effect of the innermost dusty arm shadowing the central part of the
model galaxy is determined by the scale-height of the dusty disk,
increasing from the ``thin'' through the ``standard'' to the ``thick''
model cases.  Again, the ``wide'' model extinction behaves similarly to
the ``thin'' model having the same $\chi^{\rm d}$ value.  This leads to
the conclusion that the effect of light back-scattering by dusty arms
can be neglected, while in the cases of dust distribution extending
above the stellar populations the ``shadowing'' effect becomes
significant.

\begin{figure}[H]
\centerline{\psfig{figure=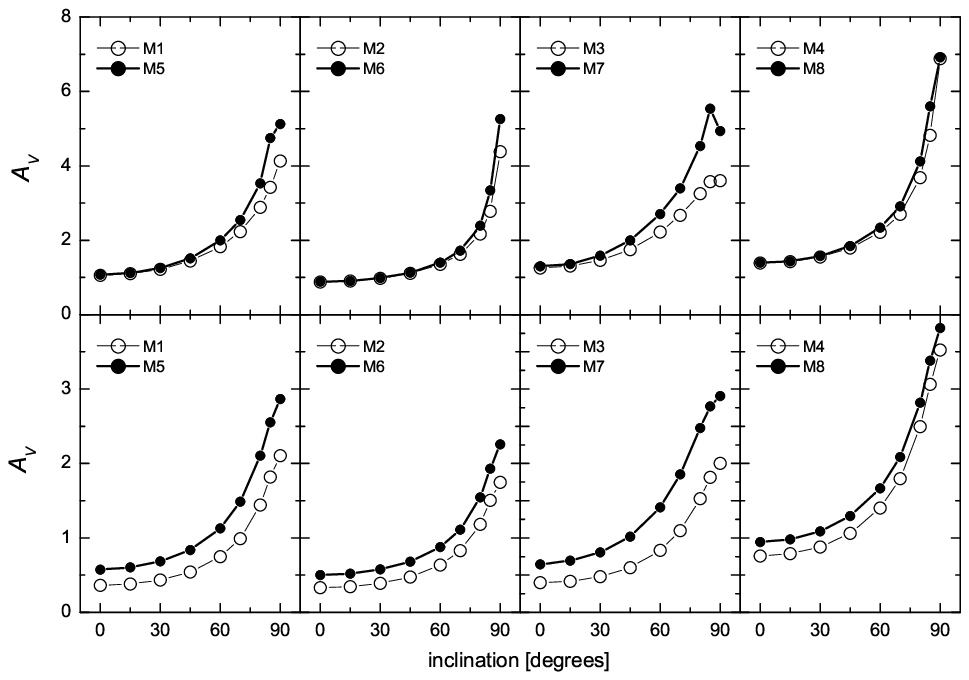,width=120mm,angle=0,clip=}}
\captionb{8}{The extinction in $V$-band, $A_V$, vs. model galaxy
inclination angle. Top row panels show $A_V$ measured in aperture
with semi-major radius of $1R_{\rm eff}$; bottom row panels show
total extinction in aperture with semi-major radius of $6R_{\rm
eff}$ (entire model galaxy). Panels are arranged in columns
according to relative dust-to-star distribution (left-to-right):
``standard'', ``thin'', ``thick'' and ``wide''. Solid lines with
filled and open circles correspond to the models with and without
dusty arms, respectively.}
\end{figure}

When considering the extinction of the entire model galaxy (Figure~8,
bottom panels) it is noticeable that the presence of dusty arms
increases the extinction by a constant value, weakly dependent on the
scale-height of the dusty disk.  This behavior is identical for all
models with inclinations between 0\degr\ and 70\degr.  The edge-on
extinction, however, is strongly dependent on the vertical extent of the
dusty arms.  This can be interpreted as dusty arms being almost a
perfect ``screen'', with scattering due to their presence, apparently
playing only a minor role in the final photometry results.  The
high-inclination models exhibit strong increase in the extinction due to
shadowing by the dusty arms, strongly dependent on the $R_{\rm eff}^{\rm
d}$.

\begin{figure}[H]
\centerline{\psfig{figure=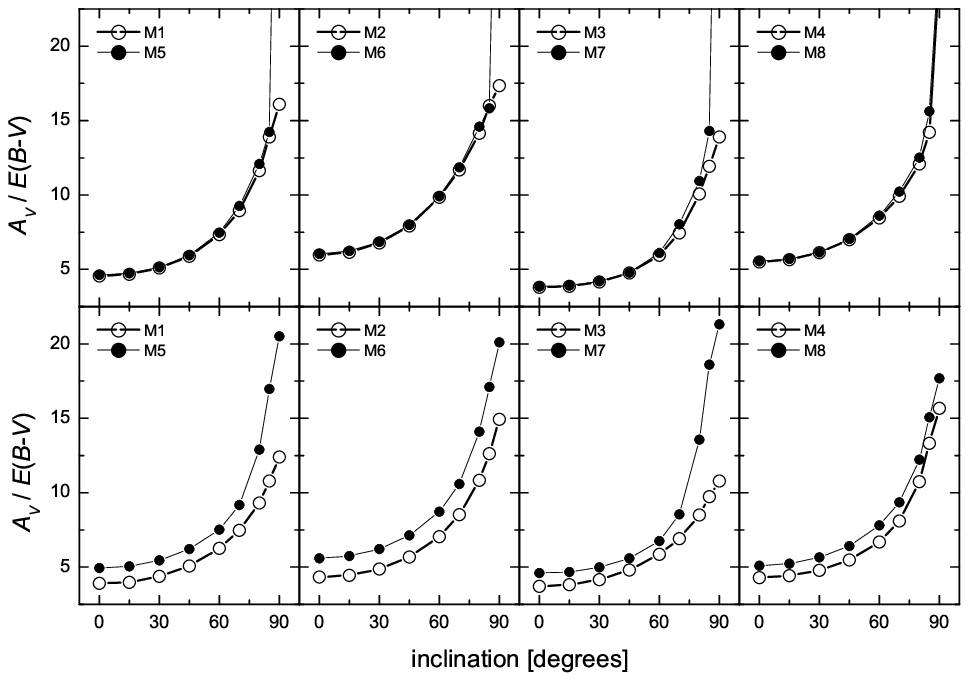,width=120mm,angle=0,clip=}}
\captionc{9}{The same as in Figure~8, but for the $A_V/E_{B-V}$
ratio.}
\end{figure}
\vspace{4mm}

\begin{figure}[H]
\centerline{\psfig{figure=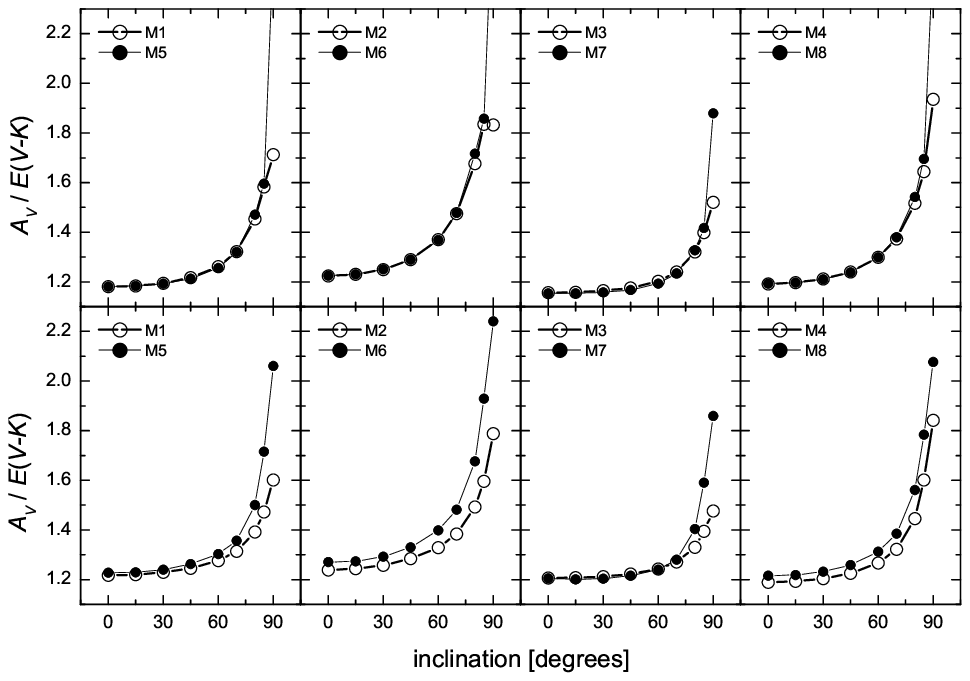,width=120mm,angle=0,clip=}}
\captionc{10}{The same as in Figure~8, but for the $A_V/E_{V-K}$
ratio.}
\end{figure}

The effects of the dusty arm presence on extinction to color excess
ratios $A_V/E_{B-V}$ and $A_V/E_{V-K}$ within aperture of $1 R_{\rm
eff}$ semi-major axis radius, shown in the upper row panels in Figures~9
and 10, respectively, are noticeable only for the ``thick'' models seen
under inclinations of 80\degr\ and higher.  This provides a direct,
although inclination-dependent, relation between color excess and
extinction for the central part of the model galaxy, which is apparently
independent of the surrounding dusty arm presence.

Extinction to color excess ratios $A_V/E_{B-V}$ and $A_V/E_{V-K}$
determined for the entire model galaxy (bottom row panels of Figures~9
and 10, respectively) show nearly constant increase for model galaxies
with dusty arms in respect to the models with a dusty disk only.  This
difference weakly depends on the scale-height of the dusty disk and
remains nearly constant up to the inclination of 70\degr\ -- for higher
inclinations it increases sharply.  This behavior can be attributed to
the increase in total dust mass of the model galaxy, and it shows almost
no effect of the dusty disk for the low inclination models.

Inclination dependencies of color excess ratio $E_{V-K}/E_{B-V}$
(Figure~11) show no effect of dusty arm presence on the photometry of
the model galaxy central part, and moderate inclination and scale-height
dependent effects for the entire model galaxy, following the trends
similar to $A_V/E_{B-V}$ ratio discussed above.  Therefore, $E_{B-V}$
and $E_{V-K}$ color excesses seem to be very good measures of total
internal extinction in disk galaxies independent of dusty arm presence.

\begin{figure}[H]
\centerline{\psfig{figure=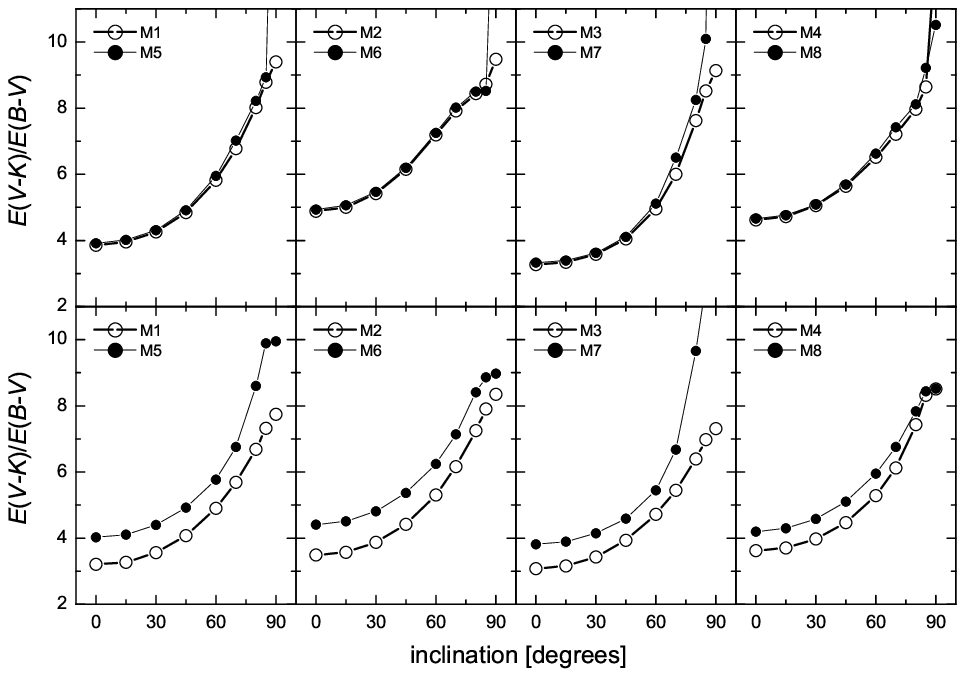,width=120mm,angle=0,clip=}}
\captionc{11}{The same as in Figure~8, but for the $E_{V-K}/E_{B-V}$
ratio.}
\end{figure}

We find no significant differences between $E_{B-V}$ and $E_{V-K}$ color
excesses of the model galaxies, having only a dusty disk and the
equivalent models with both, a dusty disk and dusty arms, possessing
smaller flatness of dust distribution than flatness of stellar
population, $\chi > \chi^{\rm d}$ (the ``thin'' and the ``wide''
models), versus the inclination of up to 70\degr.  At the same time, the
``standard'' and the ``thick'' models show significantly differing
behavior of $E_{B-V}$ and $E_{V-K}$ color excesses versus inclination
arising due to dusty arms.

\sectionb{4}{DISCUSSION AND CONCLUSIONS}

As it was demonstrated by the simplified models in section 3, dusty arms
may significantly affect the surface brightness dependence on the
inclination of disk galaxies, when the dust lanes, tightly coupled with
spiral arms, are prominent and their absorbing material extends above
the stellar disk.  On the contrary, when the dust lanes are marginal and
flattened to the central plane of the galaxy disk, their extinction
effects can be ignored as is assumed in the conventional models (see
e.g., Misiriotis \etal\ 2000).

The geometric proportions of studied model galaxies are inherent to the
real galaxy properties, see e.g., Kodaira, Watanabe \& Okamura (1986).
The pitch angle of the spiral arms is small in the early-type disk
galaxies and increases as spiral arms become loose and irregular with
increasing morphological type index.  Therefore, the ``standard'' and
the ``thin'' dusty arm models (M5 and M6) may roughly approximate the
spiral galaxies of Sa-Sc (T\,=\,1--5) types.  As the disk galaxies of
early types have a prominent bulge, which responds differently to the
extinction produced by the absorbing matter residing with the disk
(Kodaira \etal\ 1992), the shadowing effects discussed in this paper are
most relevant to the disk galaxies of the middle morphological types
Sb-Sc (T\,=\,3--5).  The special example of the Sombrero galaxy
(NGC\,4594, Sa, T\,=\,1) clearly shows an obscuring dusty ring at the
outermost part of the stellar disk, see e.g., Ohta \& Kodaira (1995).
Although, normal Sb-Sc galaxies do not show this kind of structure, the
contrast of spiral arms to the smooth stellar component in the disk
becomes stronger outwards, therefore, that also suggests the relevance
of the models, presented in this study, to the main part of normal Sb-Sc
galaxies.

While it is tempting to compare the results, obtained for the ``thick''
model to the ``boiling disk'' hypothesis proposed by Sofue \etal\
(1994), it should be done with caution.  First, the extinction
distribution in observed cases show a considerable amount of filamentary
structure, which is hard or nearly impossible to model using smooth
extinction distribution.  Second, as argued by Davies \etal\ (1998), the
blown-up dust comes from the immediate vicinity of the star-burst
regions, therefore increasing the vertical thickness of the dusty arms.
Whether this mechanism also increases the thickness of the dusty disk
remains uncertain.  The ``thick'' model was included in this study to
illustrate the extreme case opposite to the ``thin'' model and represent
the closest model to the classical ``screen'' extinction approximation.

The performed tests show that determination of the dusty disk geometry
is crucial in reconstructing the properties of global radiation field in
disk galaxies.  The fact, that the increase in the total dust mass of
the model galaxy (due to introduction of dusty arms) by over 200\% has
such a small effect on the photometric properties, leads to a conclusion
that it might be possible to model successfully photometric galaxy
parameters using stellar disks homogeneously coupled with the dusty
disks, thus supporting the current trend of ignoring the effects of
spiral arm presence and using smoothed double-exponential and similar
distributions.  At the same time, this may have a profound implication
on the efforts to determine the effective extinction using metallicity
as a single parameter of the interstellar matter mass, obtained from
observations or from evolutionary models (see e.g., Vansevi{\v c}ius,
Arimoto \& Kodaira 1997).  The results presented in this work show, that
an increase in total dust mass by a factor of 3 may change the resulting
$A_V$ extinction value by 20--100\% (even barring the extremes, seen in
our models under the inclination of $\sim$90\degr).  Therefore, in order
to estimate the effective galaxy extinction, basing on metallicity of
interstellar matter, it might be necessary to introduce a parameter
reflecting the morphology of a given galaxy in addition to the dust
clumpiness parameter suggested by Nagashima \etal\ (2002).

Both, theoretical considerations and radiative transfer models, suggest
the presence of two clearly separable effects of dusty arms:  direct
light ``shadowing'' and light ``localization'' in the regions between
dust walls.  The latter effect is more prominent and is seen in all
considered cases.  Color excesses provide a good measure of total
internal extinction for all model galaxies regardless of dusty arm
presence.  Furthermore, the results obtained with growing apertures show
that photometric properties of the model galaxies derived with large
apertures ``sense'' additional extinction produced by dusty arms.
However, the effect is much weaker than it would be in the case of the
same dust mass diffusely distributed in the dusty disk rather than in
the arms.  On the other hand, the photometry of the central part alone
is insensitive to the presence of surrounding dusty arms up to very high
inclination angles.  This may affect the conclusions drawn basing on
observations of only the central parts of distant galaxies and must be
investigated further.

The above considerations lead to the conclusion that dark lanes are
substantial contributors to the extinction effects of disk galaxies of
morphological types Sb-Sc.  We suggest that the dusty spiral arm
component should be taken into account in studies of spiral galaxy
extinction, in addition to the conventional diffuse dust layers or
disks.  However, in order to make the model galaxy study more realistic,
well adjusted specific information about the optical thickness and the
geometric parameters of spiral dark lanes, dependent on galaxy type,
must be introduced.  In future works dust clumpiness and thermal
emission effects should also be fully addressed in order to make
detailed comparison of the model spiral galaxy photometric properties
with observations.

\thanks{ We are thankful to Valdas Vansevi{\v
c}ius for correcting the manuscript. This work was financially
supported in part by a Grant T-08/06 of the Lithuanian State
Science and Studies Foundation. Computations presented in this
paper were in part performed on the computers of the Astronomical
Data Analysis Center of the National Astronomical Observatory of
Japan.}

\References
\enlargethispage{3mm}

\refb Block D. L., Elmegreen B. G., Wainscoat R. J.\ 1995, Nature, 381,
679

\refb Bosma A., Byun Y., Freeman K. C., Athanassoula E.\ 1992, ApJ, 400,
L21

\refb Burstein D., Haynes M. P., Faber S. M.\ 1991, Nature, 353, 515

\refb Corradi R.\,L.\,M., Beckman J. E., Simonneau E.\ 1996, MNRAS, 282,
1005

\refb Davies J., Alton P., Bianchi S., Trewhella M.\ 1998, MNRAS, 330,
1006

\refb Davies J., Burstein D.\ 1995, {\it The Opacity of Spiral Disks},
NATO ASI ser., Kluwer, Dordrecht

\refb de Vaucouleurs G., de Vaucouleurs A., Corwin H.\,G.\,Jr., Buta R.
J., Paturel G., Fougu P.\ 1991, {\it Third Reference Catalogue of Bright
Galaxies}, Springer-Verlag

\refb Disney M., Davies J., Phillips S.\ 1989, MNRAS, 238, 939

\refb Ferrara A., Bianchi S., Cimatti A., Giovanardi C.\ 1999, ApJS,
123, 427

\refb Holmberg E.\ 1958, Medd. Lund Obs., Ser. II, No 1, 31

\refb Huizinga J. E., van Albada T. S.\ 1992, MNRAS, 254, 677

\refb Kodaira K., Doi M., Shimasaku K.\ 1992, AJ, 104, 569

\refb Kodaira K., Watanabe M., Okamura S.\ 1986, ApJS, 62, 703

\refb Kodaira K., Yamashita T.\ 1996, PASJ, 48, 581

\refb Laor A., Draine B.\ 1993, ApJ, 402, 441

\refb Mathis J., Rumpl W., Nordsiek K.\ 1977, ApJ, 217, 425

\refb Misiriotis A., Kylafis N., Papamastorakis J., Xilouris E.\ 2000,
A\&A, 353, 117

\refb Nagashima M., Yoshii Y., Totani T., Gouda N.\ 2002, ApJ, 578, 675

\refb Ohta K., Kodaira K.\ 1995, PASJ, 47, 17

\refb Reshetnikov V., Dettmar R.-J., Combes F.\ 2003, A\&A, 399, 879

\refb Sandage A.\ 1961, {\it The Hubble Atlas of Galaxies}, Carnegie
Institution of Washington, Washington

\refb Semionov D., Vansevi{\v c}ius V.\ 2002, Baltic Astronomy, 11, 537

\refb Semionov D., Vansevi{\v c}ius V.\ 2005a, Baltic Astronomy, 14, 235

\refb Semionov D., Vansevi{\v c}ius V.\ 2005b, Baltic Astronomy, 14, 245

\refb Semionov D., Vansevi{\v c}ius V.\ 2005c, Baltic Astronomy, 14, 543

\refb Sofue Y.\ 1987, PASJ, 39, 547

\refb Sofue Y., Wakamatsu K., Malin D. F.\ 1994, AJ, 108, 2102

\refb Tomita A., Tomita Y., Saito M.\ 1996, PASJ, 48, 285

\refb Valentijin E. A.\ 1990, Nature, 346, 153

\refb Vansevi{\v c}ius V., Arimoto N., Kodaira K.\ 1997, ApJ, 474, 623

\refb Wainscoat R. J., Cohen M., Volk K., Wolker H. J., Schwartz D. E.\
1992, ApJS, 83, 111

\refb Witt A., Thornson H., Capuano J.\ 1992, ApJ, 393, 611
\end{document}